\documentclass[%
 reprint,
superscriptaddress,
 amsmath,amssymb,
 aps,
 prl,
]{revtex4-2}

\usepackage{graphicx}
\usepackage{dcolumn}
\usepackage{bm}
\usepackage{hyperref}

\usepackage[dvipsnames]{xcolor}
\usepackage{orcidlink}
\usepackage{placeins}
\usepackage{color}
\usepackage[dvipsnames]{xcolor}
\usepackage[normalem]{ulem} 

\newcommand{\Aei}{\affiliation{Max Planck Institute for Gravitational Physics (Albert Einstein Institute), D-14467 Potsdam, Germany}}
\newcommand{\Caltech}{\affiliation{Theoretical Astrophysics 350-17, California Institute of Technology, Pasadena, CA 91125, USA}}
\newcommand{\CornellPhysics}{\affiliation{Department of Physics, Cornell University, Ithaca, NY, 14853, USA}}
\newcommand{\Cornell}{\affiliation{Cornell Center for Astrophysics and Planetary Science, Cornell University, Ithaca, New York 14853, USA}}
\newcommand{\CornellLepp}{\affiliation{Laboratory for Elementary Particle Physics, Cornell University, Ithaca, New York 14853, USA}}
\newcommand{\Fullerton}{\affiliation{Nicholas and Lee Begovich Center for Gravitational-Wave Physics and Astronomy, California State University Fullerton, Fullerton, California 92831, USA}}

\newcommand{\Perimeter}{\affiliation{Perimeter Institute for Theoretical Physics, Waterloo, ON N2L2Y5, Canada}}

\newcommand{\anniversaryevent}[0]{GW250114}

\begin{document}



\author{Guillermo Lara~\orcidlink{0000-0001-9461-6292}}
\email{glara@aei.mpg.de} \Aei
\author{Harald P. Pfeiffer~\orcidlink{0000-0001-9288-519X}} \Aei

\author{Nils Deppe~\orcidlink{0000-0003-4557-4115}} \CornellLepp \CornellPhysics \Cornell
\author{Lawrence E.~Kidder~\orcidlink{0000-0001-5392-7342}} \Cornell
\author{Geoffrey Lovelace~\orcidlink{0000-0002-7084-1070}} \Fullerton
\author{Sizheng~Ma~\orcidlink{0000-0002-4645-453X}} \Perimeter
\author{Alexandra Macedo~\orcidlink{0009-0001-7671-6377}} \Fullerton
\author{Jordan Moxon~\orcidlink{0000-0001-9891-8677}} \Caltech
\author{Kyle C.~Nelli~\orcidlink{0000-0003-2426-8768}} \Caltech
\author{Mark A.~Scheel~\orcidlink{0000-0001-6656-9134}} \Caltech
\author{William Throwe~\orcidlink{0000-0001-5059-4378}} \Cornell
\author{Nils L.~Vu~\orcidlink{0000-0002-5767-3949}} \Caltech

\preprint{APS/123-QED}

\title{
    Towards long and accurate numerical relativity waveforms \\ of binary black holes beyond general relativity
}

\date{\today}

\begin{abstract}
Numerical relativity (NR) simulations of compact binaries in theories beyond general relativity (GR) will be pivotal for the continued development of future tests of gravity with gravitational waves (GWs).
In this \emph{Letter}, we show that the combination of spectral methods and the \emph{fixing-the-equations} approach allows us to produce the longest waveforms in the literature for a genuine beyond-GR theory,
thus bringing NR methods for alternative theories of gravity closer to the state-of-the-art in GR.
For concreteness, we focus on the well-known shift-symmetric version of scalar Gauss-Bonnet gravity, a theory postulating the existence of an additional dynamical scalar and describing black holes (BHs) different from the Kerr solution.
We extract the gravitational and scalar waveforms at future null infinity for equal-mass, nonspinning, eccentricity-reduced BH binaries, and quantify the phase errors to be \(\lesssim 1 \, \mathrm{rad}\) after 40+ GW cycles (20+ orbits).
We also show that the GW phase corrections in this alternative theory are distinguishable from Einstein's theory and lead to an earlier coalescence time than in GR.
Obtaining such waveforms is a stepping stone to perform precise comparisons with Post-Newtonian theory and to calibrate waveform models beyond GR.
\end{abstract}

\maketitle




\begin{figure}[]
    \includegraphics[width=\linewidth,trim=0 20 0 50,clip=True]{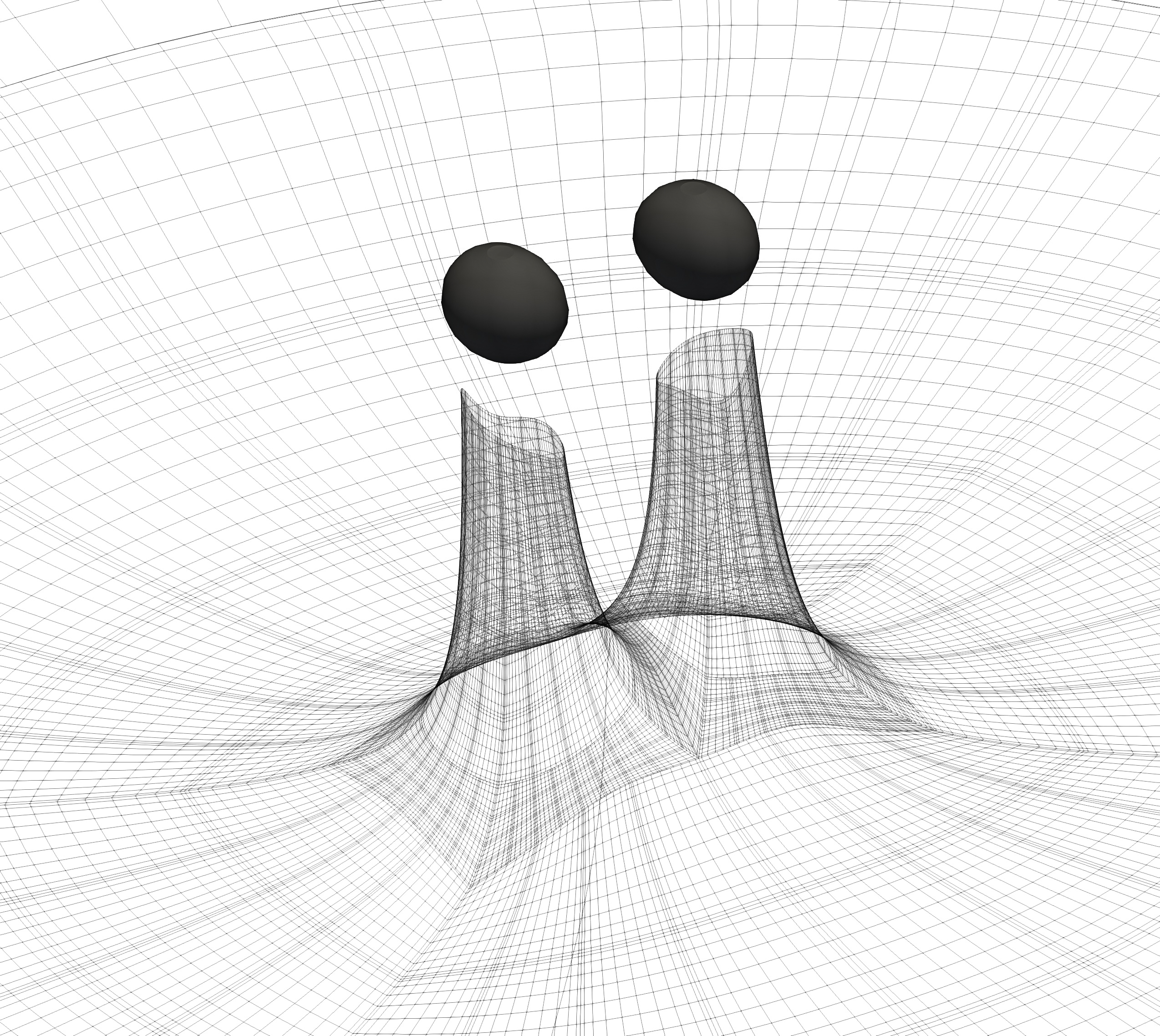}
    \caption{
        \emph{Snapshot of late-time inspiral of scalarized black holes in scalar Gauss-Bonnet gravity.} The deformed BH horizons of an equal-mass binary are shown closely before merger. The domain decomposition is illustrated with an in-plane slice of the numerical grid deformed in the vertical direction, which is proportional to the dynamical scalar \(\Psi\).
    }
    \label{fig : evolution illustration}
\end{figure}

\noindent
Gravitational wave (GW) observations of coalescing compact object binaries by the LIGO, Virgo and Kagra (LVK) observatories have significantly contributed to our knowledge of the astrophysical Universe and the nature of gravity.
In particular, tests of gravity with gravitational waves have featured prominently in analyses of events during the LVK O4 observing run~\cite{LIGOScientific:2026qni, LIGOScientific:2026fcf, LIGOScientific:2026wpt,LIGOScientific:2026uyd,
LIGOScientific:2026oim}. 
The loudest event so far, \anniversaryevent, was observed with signal-to-noise ratio (SNR) \(\sim 80\), and has allowed for unprecedented tests of the predictions of general relativity (GR) for the inspiral dynamics, remnant properties, and quasinormal mode content of the GW signal originating from an astrophysical black hole (BH) binary~\cite{LIGOScientific:2025rid, LIGOScientific:2025wao, Grimaldi:2026prn}. 
Next-generation detectors, such as the Einstein Telescope~\cite{ET:2025xjr},
Cosmic Explorer~\cite{Evans:2021gyd} and the Laser Interferometer Space Antenna~\cite{LISA:2024hlh}, will bring about observations with significantly larger SNRs, with maximum values reaching few hundreds (ground-based) and few thousands (space-based), and originating from numerous astrophysical sources.
To fully exploit the scientific potential of these observations,
substantial effort is being devoted to further develop gravitational waveform models, with numerical relativity (NR) playing a central role in providing predictions in the strong-field regime of GR.

In the same spirit, NR simulations \emph{beyond} GR can provide valuable insights for the development of future tests of gravity,
strengthening our ability to verify or falsify the predictions of GR.
For instance, NR waveforms featuring consistent deviations from GR (both in parameter space and across the entire waveform) arising from a single theory, could be used to stress-test present pipelines used to identify departures from GR.
This can be accomplished through injection studies using such waveforms, especially for scenarios that violate the assumptions in current parameterized-test implementations (e.g.~FTI~\cite{Mehta:2022pcn} and TIGER~\cite{Agathos:2013upa}).
For example, BH binaries in scalar Gauss-Bonnet (sGB) gravity that experience nonlinear ``charge-flips'' during the inspiral can abruptly introduce eccentricity during the inspiral~\cite{Lara:2025kzj}.
Besides testing existing frameworks, NR waveforms can aid construction of the waveform models themselves.
Juli\'e, \emph{et al}~\cite{Julie:2024fwy}, for instance, have built a prominent candidate for a first full waveform model beyond-GR for the case of sGB, using the effective-one-body (EOB) approach to combine information from beyond-GR computations in Post-Newtonian (PN) and BH perturbation theory.
NR waveforms obtained from a targeted simulation campaign, comprising a few nonspinning quasicircular systems, could already enable a first calibration of the model that would improve its accuracy in the nonlinear regime.
Lastly, NR waveforms can be used to perform validation studies of the predictions obtained with other computation methods ---see e.g.~Ref.~\cite{Hu:2025bkg} for a comparison of the ringdown portion of the GW signal.

Despite such numerous potential applications, NR binary BH waveforms beyond GR have been historically difficult to produce, echoing the grand challenges that characterised NR before and in the years immediately following the binary BH breakthrough~\cite{BinaryBlackHoleGrandChallengeAlliance:1997aaw, BinaryBlackHoleChallengeAlliance:1997wdh, Gomez:1998uj, Bruegmann:2003aw, Pretorius:2005gq, Baker:2005vv, Campanelli:2005dd}.
The main obstacle remains ensuring that stable numerical simulations can be achieved,
an issue closely linked to the mathematical properties of the initial (boundary-) value problem ---see e.g.~Ref.~\cite{Bernard:2019fjb} for a discussion.
Over the past decade, substantial progress has been made in this direction, leading to the emergence of a variety of new methods, applicable to broad classes of theories, as well as a number of NR codes that implement them ---see e.g.~Ref.~\cite{Ripley:2022cdh} for a review.
These include perturbative methods~\cite{Witek:2018dmd, Okounkova:2019zep, Okounkova:2019zjf, Okounkova:2020rqw}, modifications to the standard NR gauges (such as the modified generalized harmonic~\cite{Kovacs:2020pns, Kovacs:2020ywu,East:2020hgw, Corman:2022xqg}, modified CCZ4~\cite{AresteSalo:2022hua} and modified BSSN~\cite{Shum:2025lgp} systems), reformulations based on field redefinitions~\cite{Figueras:2024bba, Figueras:2025wtx}, and the \emph{fixing-the-equations} approach~\cite{Cayuso:2017iqc, Cayuso:2020lca,Lara:2021piy, Franchini:2022ukz, Cayuso:2023xbc}
---see also Refs.~\cite{Kovacs:2019jqj, Bezares:2021dma, Figueras:2021abd, Held:2023aap} for other theory-specific approaches.
However, bringing beyond-GR simulations on par with the state-of-the-art in GR, requires further improvements regarding initial data and waveforms quality ---as well as overcoming other obstacles; see e.g.~Refs.~\cite{AresteSalo:2025sxc, Thaalba:2026eyv}.

In this \emph{Letter}, we report on substantial progress in the development of the \emph{fixing-the-equations} approach~\cite{Cayuso:2017iqc} to simulate beyond-GR theories and on its numerical implementation in \textsc{spectre}~\cite{deppe_2026_19373346}, a code employing a discontinuous-Galerkin spectral scheme.
In particular, we present the longest waveforms in the literature for a genuine beyond-GR theory, i.e.\ a theory where the principal part of the equations has a different structure from that of minimally-coupled GR, opening up the possibility to perform data analysis studies and to inform waveform modelling beyond GR.
%
%
Our approach minimizes two possible sources of systematic error: \emph{(i)} the error on the intrinsic description of the component BHs, and \emph{(ii)} the error due to orbital variation of the auxiliary variables in the inertial frame.
We achieve this by introducing a new type of \emph{comoving driver} equations that exploit the approximate helical symmetry of quasicircular BH binaries.
Further details on the accuracy of our implementation are given in a companion paper~\cite{CompanionLongPaper}.


\paragraph{\textbf{Theory}.---}

\begin{figure*}
    \includegraphics[width=\textwidth]{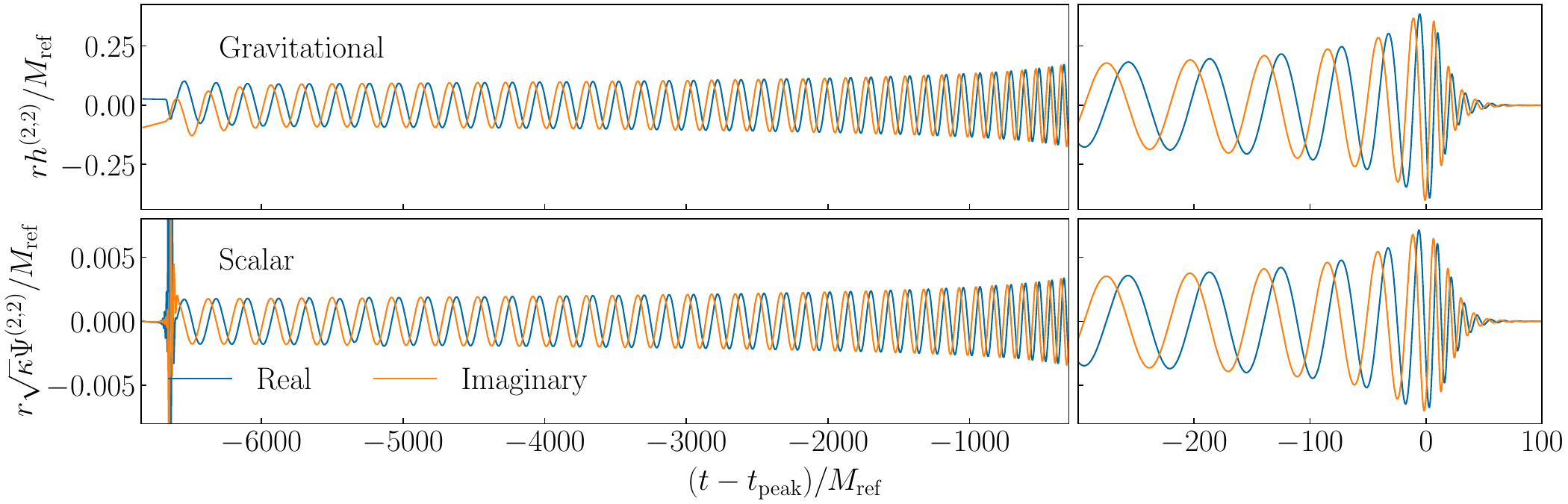}
    \caption{\emph{Gravitational and scalar waves.} 
    We show the real and imaginary parts of the \((2,2)\)-mode for the GW strain (top) and scalar field (bottom) extracted with CCE.
    Note that dipolar scalar radiation is suppressed for equal-mass nonspinning systems.
    The waves are mapped to the BMS superrest frame and the time is shown with respect to the GW peak of the \((2,2)\)-mode.
    }
    \label{fig : Showcase gw}
\end{figure*}

The action of the theory (in units where \(c=1\)) is
\begin{align} \label{eq : action}
    S\left[g_{ab}, \Psi\right] \equiv \int d^4 x \sqrt{-g} \left[\dfrac{R}{2 \kappa} - \dfrac{1}{2} \nabla_{a} \Psi \nabla^{a} \Psi  +  \ell^{2} f(\Psi) \, \mathcal{G} \right],
\end{align}
where \(\kappa = 8 \pi G \), \(g_{ab}\) is the metric [with determinant \(g \equiv \det (g_{ab})\)],  \(\Psi\) is a dynamical scalar, \(R\) is the Ricci scalar, and \(f(\Psi)\) is a shape function describing the coupling (with lengthscale \(\ell\)) to the Gauss-Bonnet scalar
\begin{align}
    \mathcal{G} &\equiv R_{abcd} R^{abcd} - 4 R_{ab} R^{ab} + R^{2} , 
\end{align}
defined in terms of contractions of the Riemann (\(R_{abcd}\)) and Ricci (\(R_{ab}\)) tensors.
We use early alphabetic letters \(\{a, b, \dots \}\) to denote spacetime indices, and \(\{i, j, \dots\}\) to denote spatial indices.
For simplicity, we focus here on the shift-symmetric case \(f(\Psi) = \Psi\), for which BHs are generically endowed with scalar \emph{hair} (see Refs.~\cite{Sotiriou:2013qea, Sotiriou:2014pfa, Capuano:2023yyh} for how this coupling to curvature evades \emph{no-hair} theorems),
and for which parameterized LVK tests have constrained to satisfy \(\sqrt{\alpha_{\mathrm{GB}}} = \ell/2 \lesssim 0.3 \, \mathrm{km}\)~\cite{Sanger:2024axs}.
The dimensionful coupling constant $\ell$ breaks the mass-invariance of vacuum GR, and our presentation below will be phrased in terms of a dimensionless coupling scale $\hat\ell^2\equiv \sqrt{\kappa}\ell^2/m^2$ normalized by the BH-mass $m$.

The action (\ref{eq : action}) yields evolution equations $R_{ab}=S_{ab}$ and $\Box\Psi=\mathcal{S}$, where the terms on the right-hand-side contain complicated combinations of second partial derivatives, which change the hyperbolic structure of the PDEs. 
Following the \emph{fixing-the-equations} approach~\cite{Cayuso:2017iqc}, 
we introduce a set of auxiliary variables \(\boldsymbol{\Sigma} \equiv (\Sigma, \Sigma_{ab})\) and obtain a new set of equations of motion,
\begin{align} \label{eq : fixed equations of motion}
    \Box \Psi &= \Sigma , &
    R_{ab} &= \Sigma_{ab} .
\end{align}
The auxiliary variables $\boldsymbol{\Sigma}$ are obtained by evolving \emph{ad hoc} driver equations that are prescribed such that the \(\boldsymbol{\Sigma}\) closely follow, or (using the terminology of Ref.~\cite{Cayuso:2023xbc}) \emph{track} the terms that include the original beyond-GR corrections \(\boldsymbol{\mathcal{S}} \equiv (\mathcal{S}, \mathcal{S}_{ab})\) in the RHS of Eqs.~\eqref{eq : fixed equations of motion}.

One of the \emph{key} insights of this \emph{Letter} is how to take advantage of any approximate (or exact) symmetries of the spacetime to construct driver equations with orders of magnitude improved tracking.
For nonspinning quasicircular BH binaries, one can identify a helical vector that describes the approximate symmetry associated to orbital motion, so that in the comoving frame of the binary, the gravitational fields appear quasi-stationary.
\emph{Comoving} coordinates \(\{\hat{t}, \hat{x}^{i}\}\)
  are related to inertial coordinates \(\{t = \hat{t}, x^{i}(\hat{t}, \hat{x}^{i})\}\)
through the \emph{frame velocity} \(v^{i} \equiv \partial x / \partial \hat{t}\).  In terms of this frame velocity, our new \emph{comoving} driver is given by
\begin{align} \label{eq : comoving driver simplified second version}
    \sigma (\partial_t + \mathcal{L}_{v})^2 \boldsymbol{\Sigma} + \tau (\partial_t + \mathcal{L}_{v}) \boldsymbol{\Sigma} = - \left(\boldsymbol{\Sigma} - \boldsymbol{\mathcal{S}}\right),
\end{align}
where \(\mathcal{L}_v\) is the Lie-derivative with respect to \(v^i\).
Here \(\{\sqrt{\sigma}, \tau\}\) are freely-specifiable positive dimensionful parameters with units of time, that control how strictly the tracking is enforced.
We generally set \(\tau = 2 \sqrt{\sigma}\) such that Eq.~\eqref{eq : comoving driver simplified second version} resembles a critically damped harmonic oscillator.

While other choices for the driver equation have been considered (including exponential equations~\cite{Allwright:2018rut}, wavelike equations~\cite{Franchini:2022ukz}, and advective equations~\cite{Cayuso:2023xbc}), the comoving driver Eq.~\eqref{eq : comoving driver simplified second version} is superior in that it recovers the solutions of the theory exactly in the stationary limit and exploits Lie-dragging to efficiently evolve the variables in \(\boldsymbol{\Sigma}\) along the orbital motion.


\paragraph{\textbf{Methodology}.---}

We adopt the viewpoint that the theory described by action~\eqref{eq : action} can be considered to be an effective field theory (EFT) and recast Eqs.~\eqref{eq : fixed equations of motion} and Eq.~\eqref{eq : comoving driver simplified second version} as a first-order system of the form
\begin{align} \label{eq : first order system}
    \partial_t \boldsymbol{u} + \mathbb{A}^{i} \left(\boldsymbol{u}\right) \partial_i \boldsymbol{u} = \boldsymbol{S}\left(\boldsymbol{u}\right),
\end{align}
where \(\mathbb{A}^{i}\left(\boldsymbol{u}\right)\) is a matrix, \(\boldsymbol{{S}}\left(\boldsymbol{u}\right)\) is a source term, and \(\boldsymbol{u}\) is a collection of first-order variables built from \(\{g_{ab}, \Psi, \Sigma, \Sigma_{ab}\}\) and their first derivatives ---see our companion paper for the details of this system.
For the metric sector, we use the first-order generalized harmonic system of Ref.~\cite{Lindblom:2005qh} and a damped harmonic gauge condition~\cite{Choptuik:2009ww, Szilagyi:2009qz, Deppe:2018uye}.
The scalar and driver sectors follow analogous first-order reductions.
Here, we extend the work of Ref.~\cite{Lara:2024rwa} and implement the tensor part of the driver systems, including all the source terms of the full equations of motion~\eqref{eq : fixed equations of motion} ---the EFT assumption is used here to perturbatively compute the source terms involving second-order time derivatives.

We use initial data in GR obtained by solving the extended conformal thin-sandwich (XCTS) equations using the elliptic solver within \textsc{spectre}~\cite{Vu:2021coj, Vu:2024cgf, Mendes:2025gov} and specify a small initial perturbation for \(\Psi\) ---see Refs.~\cite{Kovacs:2021lgk, Brady:2023dgu, Nee:2024bur} for efforts to extend initial data methods to sGB.
In the very early part of the evolution, the BH's seed scalar charge grows into its steady-state profiles around each BH.
Eccentricity reduction is carried out using an automatized algorithm designed for GR~\cite{Buonanno:2010yk}, which we find  effective reduces the eccentricity to \(e\lesssim 10^{-3}\), without requiring major changes due to the new physics of sGB gravity.
Our procedure targets low eccentricity after complete scalarization of the BHs, and therefore also eliminates changes to the orbit that would otherwise arise during scalarization of the BHs.

The evolution is carried out using the methods of Ref.~\cite{Lovelace:2024wra}, which we have extended to account for the new dynamical scalar.
We excise the interior of the BHs from the numerical domain and once the common horizon has been formed at the end of the inspiral phase, we interpolate the fields onto a new numerical domain with a single excision.

During the evolution, we record the values of the evolution variables on a set of extraction spheres and perform Cauchy Characteristic Evolution (CCE) to obtain the GW strain \(h = h_{+} - i h_{\times}\) at future null-infinity \(\mathcal{I}^{+}\)~\cite{Moxon:2021gbv}.
The scalar waves are also extracted at \(\mathcal{I}^{+}\) using a CCE extension to the Einstein-Klein-Gordon system~\cite{Ma:2024bed}.
Exploiting the residual gauge freedom at $\mathcal{I}^+$, we transform the waveforms to the Bondi-Metzner-Sachs (BMS) inspiral \emph{superrest} frame of the binary~\cite{Mitman:2021xkq, Mitman:2022kwt}.
Finally, we decompose the GW strain and the dynamical scalar into spin-weighted spherical harmonic modes, \(h = \sum_{\ell, m} h^{(\ell, m)} {_{-2} Y}_{\ell m}(\theta, \phi) \) and \(\Psi = \sum_{\ell, m} \Psi^{(\ell, m)} {Y}_{\ell m}(\theta, \phi) \), respectively.


\paragraph{\textbf{Results}.---}

We calculate quasi-circular BBH inspirals in sGB covering about 20 orbits, merger and ringdown for an equal mass binary with zero BH spin.
Figure~\ref{fig : Showcase gw} shows the real and imaginary parts of the quadrupolar $(\ell,m)=(2,2)$-mode for both the gravitational and scalar waves for a dimensionless beyond-GR coupling scale \(\hat{\ell}^2 \equiv \sqrt{\kappa}\ell^2 / m^2  = 1/20\), where \(m\) is the mass of each component BH.
The waveform is about 40 GW-cycles long and has been transformed to the inspiral BMS superrest frame.
Despite dipolar $(1,1)$ radiation being one of the main predictions of scalar tensor theories~\cite{Eardley:1975fgi, Damour:1992we} (including sGB gravity), it is suppressed for comparable-mass nonspinning systems.   Therefore, $\Psi^{(1,1)}$ vanishes for the configuration considered here, and $\Psi^{(2,2)}$ as plotted in Fig.~\ref{fig : Showcase gw} dominates.
In line with the results of Refs.~\cite{Cayuso:2017iqc, Franchini:2022ukz}, we find that successful evolutions follow the empirical relation \(\sigma \gtrsim \ell^2\).
In addition to the sGB simulations, we also perform a comparison simulation within GR by setting $\hat\ell=0$.

GW data-analysis requires high accuracy waveforms, in particular small phase errors for phase-coherent analyses like matched filtering.  Therefore, a detailed accounting of numerical errors is essential.
The numerical error of our waveforms is characterized by varying the dimensionless driver parameter \(\hat{\sigma} \equiv \sigma / m^2\) and the numerical resolution ---for the latter, we vary the number of spectral basis functions used, also known as \(p\)-refinement.
We focus on the \((2,2)\)-mode of \(h\) during the inspiral, which we decompose into amplitude and phase as \(h^{(2, 2)} = A e^{-i\Phi}\).
Given any pair of waveforms, we compute their phase difference by aligning the waveform using the \texttt{align2d} optimization algorithm~\cite{SXSPackage_v2025.0.25} on a time window covering roughly 3 orbits in the early inspiral.

\begin{figure}
    \includegraphics[width=\linewidth]{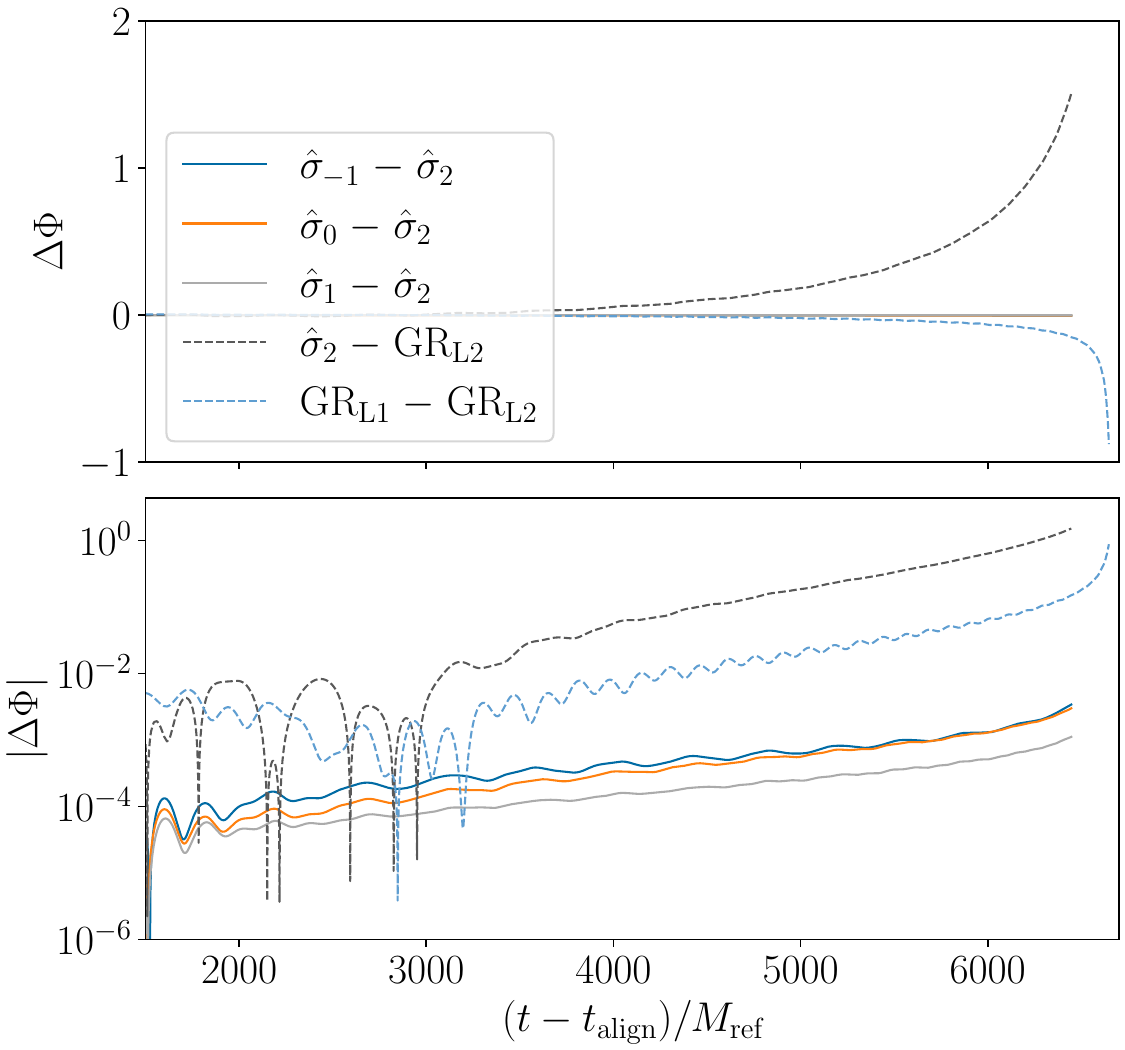}
    \caption{\emph{Dependence of the cumulative phase difference on the driver parameters.}
    The top panel shows the phase difference \(\Delta \Phi \) of the strain's \((2,2)\)-mode for driver parameters \(\hat{\sigma}_{-1} \equiv 4\), \(\hat{\sigma}_0 \equiv 1 \), \(\hat{\sigma}_1 \equiv 1/4 \), \(\hat{\sigma}_2 \equiv 1/16 \).
    The resolution is fixed to L1 (middle resolution), unless otherwise stated.
    The phase difference with respect to GR (dashed black) is distinguishable from the phase error due to varying \(\hat{\sigma}\), as well as numerical resolution, here shown for the GR reference simulation (dashed blue).
    The waveforms are aligned on a time window near \((t-t^{}_{\mathrm{peak}}) /  (M_\mathrm{ref} ) \sim -5000 \) in Fig.~\ref{fig : Showcase gw}.
    In the bottom panel, we show the magnitude of \(\Delta \Phi\).
    }
    \label{fig : gw dephasing sigma}
\end{figure}

In Fig.~\ref{fig : gw dephasing sigma}, we show the difference in phase evolution \(\Delta \Phi \) between waveforms obtained with different values of \(\hat{\sigma}\), and with respect to a simulation in GR, for which we set \(\hat{\ell}^2 = 0\).
The phase difference is shown up to the waveform peak for the difference between the two GR resolutions, or until the simulation stops.
For the \(\hat{\sigma}_{2} \equiv 1/16\) the simulation could not be continued past merger, either because \(\sigma^2\) is too close to \(\ell^2\), or because we have not found for this case an optimal setup for the transition to ringdown; the inspiral portion of this simulation should however suffice to quantify the accuracy of the method.
Changing \(\hat{\sigma}\) results in an accumulated phase error that is at most \(\sim 10^{-3} \, \mathrm{rad}\) during the inspiral and just before merger, out of an accumulated phase of $\approx 250\,\mathrm{rad}$.
Numerical truncation error is shown by the line ${\rm GR}_{\rm L1}-{\rm GR}_{\rm L2}$ in Fig.~\ref{fig : gw dephasing sigma}; this is larger than the sGB approximation errors by 2 orders of magnitude, illustrating how well our new driver conditions work.
Truncation error for the sGB simulations is comparable.

Finally, the physically relevant phase-difference between sGB and GR (dashed black) is another order of magnitude larger than the numerical truncation error, indicating that our simulations reliably resolve the physical effects over the entire 20 orbit inspiral, and making the GR and beyond-GR waveforms \emph{distinguishable} from each other.

\begin{figure}
    \includegraphics[width=\linewidth]{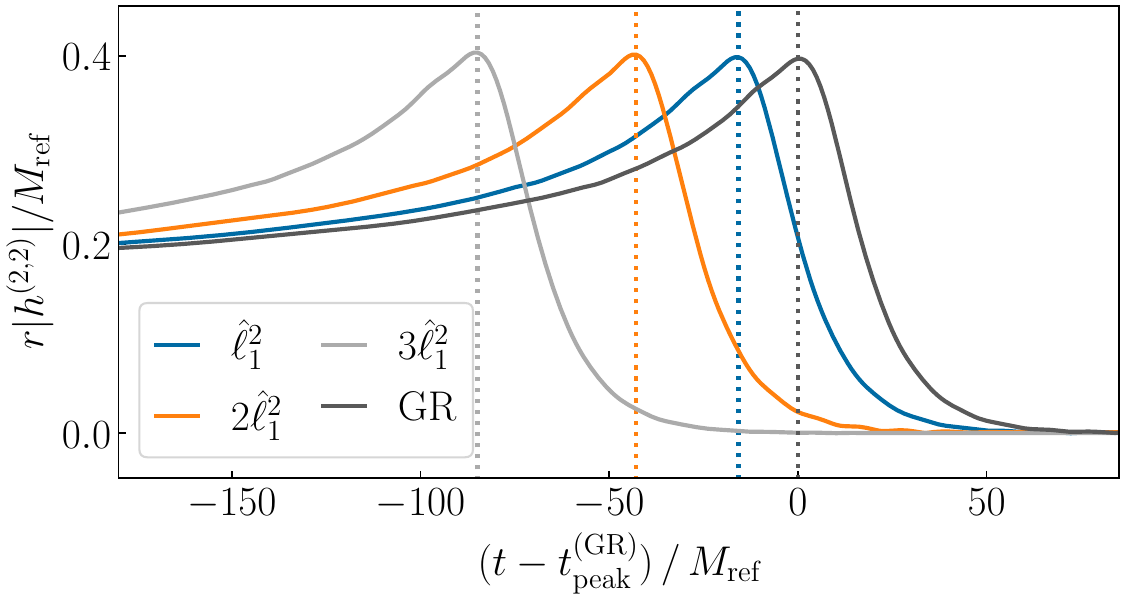}
    \caption{\emph{Merger time comparison for different couplings.} 
    The waveform amplitudes are shown for different multiples of the beyond-GR coupling \(\hat{\ell}^{2}_1 = 1/40\) aligned in the early inspiral to a GR evolution (\(\hat{\ell}^2 = 0\)).
    We keep the numerical resolution fixed to L1 (our middle resolution).
    }
    \label{fig : merger time difference with GR couplings}
\end{figure}

Having quantified the accuracy of our evolutions (for further checks see our companion paper~\cite{CompanionLongPaper}), we can probe the beyond-GR corrections to the dynamics of binaries in sGB gravity.
In particular, we address the effect of the corrections to coalescence time for equal-mass nonspinning systems.
Besides the nonlinear aspect of the problem, several factors already affect the dynamics in the weak field, including:
\emph{(i)} more radiated energy than in GR in the form of scalar waves;
\emph{(ii)} an additional attractive long-range scalar force between the objects (often written in PN theory as a rescaling of the ``bare'' gravitational constant \(G \to (1 + \alpha_{A} \alpha_{B}) G \), where \(\alpha_{A,B}\) are proportional to the respective scalar charges \(Q_{A,B}\) of the component BHs).
For generic binaries, the dipolar energy flux is enhanced by a factor \(D^{-1}\) at large separations \(D\) with respect to the quadrupolar GW emission predicted in GR (i.e.~it contributes at \(-1\)PN-order in the energy flux).
Provided that the systems are compared from sufficiently large separations, the beyond-GR corrections would then lead to an earlier coalescence ---one would expect that \emph{(i)} accelerates circularization; see e.g.~Ref.~\cite{Cardoso:2020iji}.
In the equal-mass case, however, the dipolar radiation vanishes and the beyond-GR corrections to the energy flux are no longer enhanced at large separations (entering at the same PN order as the GW flux in GR).
One may thus worry about a subtle interplay between \emph{(i)} and \emph{(ii)}, especially in the nonlinear regime: while \emph{(i)} contributes to radiated energy, \emph{(ii)} increases the amount of binding energy that needs to be radiated.
In the following, we show how NR simulations may shed light on this issue.
In Fig.~\ref{fig : merger time difference with GR couplings}, we show the amplitude of the GW strain for different values of \(\hat{\ell}^2\), including GR, for waveforms that are aligned in the early inspiral as above.
The location of the peak of the waveform (dashed lines), associated to merger, shifts earlier in time as the \(\hat{\ell}^2\) is increased.
This leads us to conclude that the beyond-GR corrections result in an earlier coalescence than in GR, at odds with the original results \footnote{
    We have been in correspondence with the authors, who have confirmed that their updated results no longer indicate a delayed merger for the beyond-GR case, and which they will present in a revised version of the paper.
}
of Ref.~\cite{Corman:2025wun}.


\paragraph{\textbf{Conclusions}.---}


GW astronomy is steadily turning into a data-rich field, creating new opportunities to harness observational data to improve our understanding of gravity.
In preparation for observations with next-generation GW detectors, efforts are underway to further develop both waveform models and tests of GR.
In this \emph{Letter}, we have shown the capability of \textsc{spectre} to produce long and accurate waveforms of the type that are needed to allow injection studies covering low masses, precise comparisons to PN theory, and calibration of waveform models beyond-GR, and more broadly, to enable better forecasts for tests of GR.
For concreteness, we focused on the case of shift-symmetric sGB gravity.
However, we can easily adjust our implementation to explore other variants of sGB gravity and in principle the \emph{fixing-the-equations} approach is flexible enough to accommodate other theories (e.g.~EFT extensions of vacuum GR~\cite{Cayuso:2020lca, Cayuso:2023xbc}, \(k\)-essence~\cite{Bezares:2021yek, Lara:2021piy, Coates:2023swo}, self-interacting massive vectors~\cite{Rubio:2024ryv}).

A \emph{key} advantage of \textsc{spectre} is that it incorporates several state-of-the-art NR methods originally developed for GR, and which we have been systematically extending to alternative theories of gravity, including:
\emph{(i)} spectral methods are accurate and  efficient for long inspirals;
\emph{(ii)} quasistationary initial data enables targeting particular systems and reducing transients;
\emph{(iii)} eccentricity control algorithms~\cite{Buonanno:2010yk}, well-tested in \textsc{SpEC}~\cite{Mroue:2013xna, Boyle:2019kee, Scheel:2025jct}, can be ported over and adjusted;
\emph{(iv)} advanced waveform extraction methods (Cauchy Characteristic Evolution~\cite{Moxon:2021gbv, Ma:2024bed} and BMS frame fixing~\cite{Mitman:2021xkq, Mitman:2022kwt}), for \emph{both} the gravitational and scalar waves, allow us to obtain accurate waveforms ---naturally including \emph{memory} (see our companion paper~\cite{CompanionLongPaper} and also Refs.~\cite{Gasparotto:2026bru, Zosso:2026uty}).
All of these will prove useful in overcoming current obstacles to making precise predictions for the strong-field dynamics beyond-GR, and in systematically exploring the parameter spaces of theories that capture representative features of possible beyond-GR phenomena.

The current situation, in which several NR codes are capable of simulating the same beyond-GR systems (using different formulations of the equations and/or numerical methods), is a fertile ground for cross-code comparisons.
Given the complexity of the relativistic two-body problem, such redundancy is essential for ensuring the reliability of any beyond-GR predictions that make their way into future tests of GR.
In moving forward, the community may draw inspiration from comparisons started more than two decades ago for predictions in GR~\cite{Alcubierre:2003pc, Babiuc:2007vr, Hannam:2009hh,Aylott:2009ya, Aylott:2009tn}.


\paragraph{\textbf{Acknowledgments}.---}
    The authors would like to thank Alessandra Buonanno, Fabrizio Corelli, Luis Lehner, Gonzalo Morr\'as, Oliver Markwell, Peter James Nee, Raj Patil, and Sebastian V\"olkel for fruitful discussions.
    Computations were performed on the Urania HPC systems at the Max Planck Computing and Data Facility.
    This material is based upon work supported by the National Science Foundation under Grants No.~PHY-2309211; No.~PHY-2309231;  No.~OAC-2513339 at Caltech; and NASA award No.~80NSSC26K0340, and No.~PHY-2407742; No.~PHY-2207342; No.~OAC-2513338; and NASA award No.~80NSSC26K0340 at Cornell; NSF Grants NSF AST-2219109, PHY-2208014, Nicholas and Lee Begovich, and the Dan Black Family Trust. Any opinions, findings, and conclusions or recommendations expressed in this material are those of the author(s) and do not necessarily reflect the views of the National Science Foundation or NASA. This work was supported by the Sherman Fairchild Foundation at Caltech and Cornell.
    NLV acknowledges support from the Swiss National Science Foundation (SNSF) Ambizione grant PZ00-2\_232961.

\bibliography{paper}

@unpublished{Evans:2021gyd,
    author = "Evans, Matthew and others",
    title = "{A Horizon Study for Cosmic Explorer: Science, Observatories, and Community}",
    eprint = "2109.09882",
    archivePrefix = "arXiv",
    primaryClass = "astro-ph.IM",
    reportNumber = "CE-P2100003-v7, Cosmic Explorer technical report CE-P2100003-v6",
    month = "9",
    year = "2021"
}

@article{Lara:2024rwa,
    author = "Lara, Guillermo and Pfeiffer, Harald P. and Wittek, Nikolas A. and Vu, Nils L. and Nelli, Kyle C. and Carpenter, Alexander and Lovelace, Geoffrey and Scheel, Mark A. and Throwe, William",
    title = "{Scalarization of isolated black holes in scalar Gauss-Bonnet theory in the fixing-the-equations approach}",
    eprint = "2403.08705",
    archivePrefix = "arXiv",
    primaryClass = "gr-qc",
    doi = "10.1103/PhysRevD.110.024033",
    journal = "Phys. Rev. D",
    volume = "110",
    number = "2",
    pages = "024033",
    year = "2024"
}

@article{Lindblom:2005qh,
    author = "Lindblom, Lee and Scheel, Mark A. and Kidder, Lawrence E. and Owen, Robert and Rinne, Oliver",
    title = "{A New generalized harmonic evolution system}",
    eprint = "gr-qc/0512093",
    archivePrefix = "arXiv",
    doi = "10.1088/0264-9381/23/16/S09",
    journal = "Class. Quant. Grav.",
    volume = "23",
    pages = "S447--S462",
    year = "2006"
}

@article{Nee:2024bur,
    author = "Nee, Peter James and Lara, Guillermo and Pfeiffer, Harald P. and Vu, Nils L.",
    title = "{Quasistationary hair for binary black hole initial data in scalar Gauss-Bonnet gravity}",
    eprint = "2406.08410",
    archivePrefix = "arXiv",
    primaryClass = "gr-qc",
    doi = "10.1103/PhysRevD.111.024061",
    journal = "Phys. Rev. D",
    volume = "111",
    number = "2",
    pages = "024061",
    year = "2025"
}

@article{Mendes:2025gov,
    author = "Mendes, Iago B. and Vu, Nils L. and Long, Oliver and Pfeiffer, Harald P. and Owen, Robert",
    title = "{Parameter control for binary black hole initial data}",
    eprint = "2509.07291",
    archivePrefix = "arXiv",
    primaryClass = "gr-qc",
    doi = "10.1103/zh31-bbtm",
    journal = "Phys. Rev. D",
    volume = "112",
    number = "12",
    pages = "124049",
    year = "2025"
}

@article{Lovelace:2024wra,
    author = "Lovelace, Geoffrey and others",
    title = "{Simulating binary black hole mergers using discontinuous Galerkin methods}",
    eprint = "2410.00265",
    archivePrefix = "arXiv",
    primaryClass = "gr-qc",
    doi = "10.1088/1361-6382/ad9f19",
    journal = "Class. Quant. Grav.",
    volume = "42",
    number = "3",
    pages = "035001",
    year = "2025"
}

@article{Damour:1992we,
    author = "Damour, Thibault and Esposito-Farese, Gilles",
    title = "{Tensor multiscalar theories of gravitation}",
    reportNumber = "IHES-P-91-93, CPT-91-PE-2542",
    doi = "10.1088/0264-9381/9/9/015",
    journal = "Class. Quant. Grav.",
    volume = "9",
    pages = "2093--2176",
    year = "1992"
}

@article{Okounkova:2019zjf,
    author = "Okounkova, Maria and Stein, Leo C. and Moxon, Jordan and Scheel, Mark A. and Teukolsky, Saul A.",
    title = "{Numerical relativity simulation of GW150914 beyond general relativity}",
    eprint = "1911.02588",
    archivePrefix = "arXiv",
    primaryClass = "gr-qc",
    doi = "10.1103/PhysRevD.101.104016",
    journal = "Phys. Rev. D",
    volume = "101",
    number = "10",
    pages = "104016",
    year = "2020"
}

@article{Brady:2023dgu,
    author = "Brady, Sam E. and Arest{\'e} Sal{\'o}, Llibert and Clough, Katy and Figueras, Pau and S., Annamalai P.",
    title = "{Solving the initial conditions problem for modified gravity theories}",
    eprint = "2308.16791",
    archivePrefix = "arXiv",
    primaryClass = "gr-qc",
    doi = "10.1103/PhysRevD.108.104022",
    journal = "Phys. Rev. D",
    volume = "108",
    number = "10",
    pages = "104022",
    year = "2023"
}

@article{Sotiriou:2013qea,
    author = "Sotiriou, Thomas P. and Zhou, Shuang-Yong",
    title = "{Black hole hair in generalized scalar-tensor gravity}",
    eprint = "1312.3622",
    archivePrefix = "arXiv",
    primaryClass = "gr-qc",
    doi = "10.1103/PhysRevLett.112.251102",
    journal = "Phys. Rev. Lett.",
    volume = "112",
    pages = "251102",
    year = "2014"
}

@article{Sotiriou:2014pfa,
    author = "Sotiriou, Thomas P. and Zhou, Shuang-Yong",
    title = "{Black hole hair in generalized scalar-tensor gravity: An explicit example}",
    eprint = "1408.1698",
    archivePrefix = "arXiv",
    primaryClass = "gr-qc",
    doi = "10.1103/PhysRevD.90.124063",
    journal = "Phys. Rev. D",
    volume = "90",
    pages = "124063",
    year = "2014"
}

@article{Cayuso:2017iqc,
    author = "Cayuso, Juan and Ortiz, N{\'e}stor and Lehner, Luis",
    title = "{Fixing extensions to general relativity in the nonlinear regime}",
    eprint = "1706.07421",
    archivePrefix = "arXiv",
    primaryClass = "gr-qc",
    doi = "10.1103/PhysRevD.96.084043",
    journal = "Phys. Rev. D",
    volume = "96",
    number = "8",
    pages = "084043",
    year = "2017"
}

@article{Lara:2021piy,
    author = "Lara, Guillermo and Bezares, Miguel and Barausse, Enrico",
    title = "{UV completions, fixing the equations, and nonlinearities in k-essence}",
    eprint = "2112.09186",
    archivePrefix = "arXiv",
    primaryClass = "gr-qc",
    doi = "10.1103/PhysRevD.105.064058",
    journal = "Phys. Rev. D",
    volume = "105",
    number = "6",
    pages = "064058",
    year = "2022"
}

@article{Vu:2021coj,
    author = "Vu, Nils L. and others",
    title = "{A scalable elliptic solver with task-based parallelism for the SpECTRE numerical relativity code}",
    eprint = "2111.06767",
    archivePrefix = "arXiv",
    primaryClass = "gr-qc",
    doi = "10.1103/PhysRevD.105.084027",
    journal = "Phys. Rev. D",
    volume = "105",
    number = "8",
    pages = "084027",
    year = "2022"
}

@article{Vu:2024cgf,
    author = "Vu, Nils L.",
    title = "{Discontinuous Galerkin scheme for elliptic equations on extremely stretched grids}",
    eprint = "2405.06120",
    archivePrefix = "arXiv",
    primaryClass = "gr-qc",
    doi = "10.1103/PhysRevD.110.084062",
    journal = "Phys. Rev. D",
    volume = "110",
    number = "8",
    pages = "084062",
    year = "2024"
}

@article{Kovacs:2021lgk,
    author = "Kovacs, Aron D.",
    title = "{On the construction of asymptotically flat initial data in scalar-tensor effective field theory}",
    eprint = "2103.06895",
    archivePrefix = "arXiv",
    primaryClass = "gr-qc",
    month = "3",
    year = "2021",
    journal = "",
}

@article{Scheel:2025jct,
    author = "Scheel, Mark A. and others",
    title = "{The SXS collaboration{\textquoteright}s third catalog of binary black hole simulations}",
    eprint = "2505.13378",
    archivePrefix = "arXiv",
    primaryClass = "gr-qc",
    doi = "10.1088/1361-6382/adfd34",
    journal = "Class. Quant. Grav.",
    volume = "42",
    number = "19",
    pages = "195017",
    year = "2025"
}

@article{Moxon:2021gbv,
    author = "Moxon, Jordan and Scheel, Mark A. and Teukolsky, Saul A. and Deppe, Nils and Fischer, Nils and H{\'e}bert, Francois and Kidder, Lawrence E. and Throwe, William",
    title = "{SpECTRE Cauchy-characteristic evolution system for rapid, precise waveform extraction}",
    eprint = "2110.08635",
    archivePrefix = "arXiv",
    primaryClass = "gr-qc",
    doi = "10.1103/PhysRevD.107.064013",
    journal = "Phys. Rev. D",
    volume = "107",
    number = "6",
    pages = "064013",
    year = "2023"
}

@article{Ma:2024bed,
    author = "Ma, Sizheng and Nelli, Kyle C. and Moxon, Jordan and Scheel, Mark A. and Deppe, Nils and Kidder, Lawrence E. and Throwe, William and Vu, Nils L.",
    title = "{Einstein{\textendash}Klein{\textendash}Gordon system via Cauchy-characteristic evolution: computation of memory and ringdown tail}",
    eprint = "2409.06141",
    archivePrefix = "arXiv",
    primaryClass = "gr-qc",
    doi = "10.1088/1361-6382/adaf6f",
    journal = "Class. Quant. Grav.",
    volume = "42",
    number = "5",
    pages = "055006",
    year = "2025"
}

@article{East:2020hgw,
    author = "East, William E. and Ripley, Justin L.",
    title = "{Evolution of Einstein-scalar-Gauss-Bonnet gravity using a modified harmonic formulation}",
    eprint = "2011.03547",
    archivePrefix = "arXiv",
    primaryClass = "gr-qc",
    doi = "10.1103/PhysRevD.103.044040",
    journal = "Phys. Rev. D",
    volume = "103",
    number = "4",
    pages = "044040",
    year = "2021"
}

@article{Bernard:2019fjb,
    author = "Bernard, Laura and Lehner, Luis and Luna, Raimon",
    title = "{Challenges to global solutions in Horndeski{\textquoteright}s theory}",
    eprint = "1904.12866",
    archivePrefix = "arXiv",
    primaryClass = "gr-qc",
    doi = "10.1103/PhysRevD.100.024011",
    journal = "Phys. Rev. D",
    volume = "100",
    number = "2",
    pages = "024011",
    year = "2019"
}

@article{Franchini:2022ukz,
    author = "Franchini, Nicola and Bezares, Miguel and Barausse, Enrico and Lehner, Luis",
    title = "{Fixing the dynamical evolution in scalar-Gauss-Bonnet gravity}",
    eprint = "2206.00014",
    archivePrefix = "arXiv",
    primaryClass = "gr-qc",
    doi = "10.1103/PhysRevD.106.064061",
    journal = "Phys. Rev. D",
    volume = "106",
    number = "6",
    pages = "064061",
    year = "2022"
}

@article{Lara:2025kzj,
    author = "Lara, Guillermo and others",
    title = "{Signatures from metastable oppositely-charged black hole binaries in scalar Gauss-Bonnet gravity}",
    eprint = "2505.14785",
    archivePrefix = "arXiv",
    primaryClass = "gr-qc",
    month = "5",
    year = "2025",
    journal = "",
}

@article{Hu:2025bkg,
    author = "Hu, Zexin and Doneva, Daniela D. and Yazadjiev, Stoytcho S. and Shao, Lijing",
    title = "{Quasinormal mode ringing of binary black hole mergers in scalar-Gauss-Bonnet gravity}",
    eprint = "2511.20301",
    archivePrefix = "arXiv",
    primaryClass = "gr-qc",
    doi = "10.1103/dtd2-5vlg",
    journal = "Phys. Rev. D",
    volume = "113",
    number = "4",
    pages = "044041",
    year = "2026"
}

@article{Shum:2025lgp,
    author = "Shum, Harry L. H. and Arest{\'e} Sal{\'o}, Llibert and Thaalba, Farid and Bezares, Miguel and Sotiriou, Thomas P.",
    title = "{A well-posed BSSN-type formulation for scalar-tensor theories of gravity with second-order field equations}",
    eprint = "2512.11034",
    archivePrefix = "arXiv",
    primaryClass = "gr-qc",
    month = "12",
    year = "2025",
    journal = "",
}

@article{Kovacs:2020pns,
    author = "Kov{\'a}cs, {\'A}ron D. and Reall, Harvey S.",
    title = "{Well-Posed Formulation of Scalar-Tensor Effective Field Theory}",
    eprint = "2003.04327",
    archivePrefix = "arXiv",
    primaryClass = "gr-qc",
    doi = "10.1103/PhysRevLett.124.221101",
    journal = "Phys. Rev. Lett.",
    volume = "124",
    number = "22",
    pages = "221101",
    year = "2020"
}

@article{Kovacs:2020ywu,
    author = "Kov{\'a}cs, {\'A}ron D. and Reall, Harvey S.",
    title = "{Well-posed formulation of Lovelock and Horndeski theories}",
    eprint = "2003.08398",
    archivePrefix = "arXiv",
    primaryClass = "gr-qc",
    doi = "10.1103/PhysRevD.101.124003",
    journal = "Phys. Rev. D",
    volume = "101",
    number = "12",
    pages = "124003",
    year = "2020"
}

@article{Figueras:2024bba,
    author = "Figueras, Pau and Held, Aaron and Kov{\'a}cs, {\'A}ron D.",
    title = "{Well-posed initial value formulation of general effective field theories of gravity}",
    eprint = "2407.08775",
    archivePrefix = "arXiv",
    primaryClass = "gr-qc",
    month = "7",
    year = "2024",
    journal = "",
}

@article{Cayuso:2023xbc,
    author = "Cayuso, Ramiro and Figueras, Pau and Fran{\c{c}}a, Tiago and Lehner, Luis",
    title = "{Self-Consistent Modeling of Gravitational Theories beyond General Relativity}",
    eprint = "2303.07246",
    archivePrefix = "arXiv",
    primaryClass = "gr-qc",
    doi = "10.1103/PhysRevLett.131.111403",
    journal = "Phys. Rev. Lett.",
    volume = "131",
    number = "11",
    pages = "111403",
    year = "2023"
}

@article{Cayuso:2020lca,
    author = "Cayuso, Ramiro and Lehner, Luis",
    title = "{Nonlinear, noniterative treatment of EFT-motivated gravity}",
    eprint = "2005.13720",
    archivePrefix = "arXiv",
    primaryClass = "gr-qc",
    doi = "10.1103/PhysRevD.102.084008",
    journal = "Phys. Rev. D",
    volume = "102",
    number = "8",
    pages = "084008",
    year = "2020"
}

@article{Figueras:2025wtx,
    author = "Figueras, Pau and Kov{\'a}cs, {\'A}ron D. and Yao, Shunhui",
    title = "{Stable non-linear evolution in regularised higher derivative effective field theories}",
    eprint = "2505.00082",
    archivePrefix = "arXiv",
    primaryClass = "hep-th",
    reportNumber = "EFI-25-4",
    doi = "10.1007/JHEP10(2025)150",
    journal = "JHEP",
    volume = "10",
    pages = "150",
    year = "2025"
}

@article{Corman:2022xqg,
    author = "Corman, Maxence and Ripley, Justin L. and East, William E.",
    title = "{Nonlinear studies of binary black hole mergers in Einstein-scalar-Gauss-Bonnet gravity}",
    eprint = "2210.09235",
    archivePrefix = "arXiv",
    primaryClass = "gr-qc",
    doi = "10.1103/PhysRevD.107.024014",
    journal = "Phys. Rev. D",
    volume = "107",
    number = "2",
    pages = "024014",
    year = "2023"
}

@article{Witek:2018dmd,
    author = "Witek, Helvi and Gualtieri, Leonardo and Pani, Paolo and Sotiriou, Thomas P.",
    title = "{Black holes and binary mergers in scalar Gauss-Bonnet gravity: scalar field dynamics}",
    eprint = "1810.05177",
    archivePrefix = "arXiv",
    primaryClass = "gr-qc",
    doi = "10.1103/PhysRevD.99.064035",
    journal = "Phys. Rev. D",
    volume = "99",
    number = "6",
    pages = "064035",
    year = "2019"
}

@article{Okounkova:2019zep,
    author = "Okounkova, Maria",
    title = "{Stability of Rotating Black Holes in Einstein Dilaton Gauss-Bonnet Gravity}",
    eprint = "1909.12251",
    archivePrefix = "arXiv",
    primaryClass = "gr-qc",
    doi = "10.1103/PhysRevD.100.124054",
    journal = "Phys. Rev. D",
    volume = "100",
    number = "12",
    pages = "124054",
    year = "2019"
}

@article{Okounkova:2020rqw,
    author = "Okounkova, Maria",
    title = "{Numerical relativity simulation of GW150914 in Einstein dilaton Gauss-Bonnet gravity}",
    eprint = "2001.03571",
    archivePrefix = "arXiv",
    primaryClass = "gr-qc",
    doi = "10.1103/PhysRevD.102.084046",
    journal = "Phys. Rev. D",
    volume = "102",
    number = "8",
    pages = "084046",
    year = "2020"
}

@article{LIGOScientific:2025rid,
    author = "Abac, A. G. and others",
    collaboration = "LIGO Scientific, Virgo, KAGRA",
    title = "{GW250114: Testing Hawking's Area Law and the Kerr Nature of Black Holes}",
    eprint = "2509.08054",
    archivePrefix = "arXiv",
    primaryClass = "gr-qc",
    reportNumber = "LIGO-P2500421",
    doi = "10.1103/kw5g-d732",
    journal = "Phys. Rev. Lett.",
    volume = "135",
    number = "11",
    pages = "111403",
    year = "2025"
}

@article{LIGOScientific:2025wao,
    author = "Abac, A. G. and others",
    collaboration = "LIGO Scientific, Virgo, KAGRA",
    title = "{Black Hole Spectroscopy and Tests of General Relativity with GW250114}",
    eprint = "2509.08099",
    archivePrefix = "arXiv",
    primaryClass = "gr-qc",
    reportNumber = "LIGO P2500461",
    doi = "10.1103/6c61-fm1n",
    journal = "Phys. Rev. Lett.",
    volume = "136",
    number = "4",
    pages = "041403",
    year = "2026"
}

@article{Grimaldi:2026prn,
    author = "Grimaldi, Leonardo and Maggio, Elisa and Pompili, Lorenzo and Buonanno, Alessandra",
    title = "{Plunge-Merger-Ringdown Tests of General Relativity with GW250114}",
    eprint = "2601.13173",
    archivePrefix = "arXiv",
    primaryClass = "gr-qc",
    month = "1",
    year = "2026",
    journal = "",
}

@article{Boyle:2019kee,
    author = "Boyle, Michael and others",
    title = "{The SXS Collaboration catalog of binary black hole simulations}",
    eprint = "1904.04831",
    archivePrefix = "arXiv",
    primaryClass = "gr-qc",
    doi = "10.1088/1361-6382/ab34e2",
    journal = "Class. Quant. Grav.",
    volume = "36",
    number = "19",
    pages = "195006",
    year = "2019"
}

@article{LIGOScientific:2026qni,
    author = "Abac, A. G. and others",
    collaboration = "LIGO Scientific, VIRGO, KAGRA",
    title = "{GWTC-4.0: Tests of General Relativity. I. Overview and General Tests}",
    eprint = "2603.19019",
    archivePrefix = "arXiv",
    primaryClass = "gr-qc",
    reportNumber = "LIGO-P2500065",
    month = "3",
    year = "2026",
    journal = "",
}

@article{LIGOScientific:2026fcf,
    author = "Abac, A. G. and others",
    collaboration = "LIGO Scientific, VIRGO, KAGRA",
    title = "{GWTC-4.0: Tests of General Relativity. II. Parameterized Tests}",
    eprint = "2603.19020",
    archivePrefix = "arXiv",
    primaryClass = "gr-qc",
    reportNumber = "LIGO-P2500066",
    month = "3",
    year = "2026",
    journal = "",
}

@article{LIGOScientific:2026wpt,
    author = "Abac, A. G. and others",
    collaboration = "LIGO Scientific, VIRGO, KAGRA",
    title = "{GWTC-4.0: Tests of General Relativity. III. Tests of the Remnants}",
    eprint = "2603.19021",
    archivePrefix = "arXiv",
    primaryClass = "gr-qc",
    reportNumber = "LIGO-P2500067",
    month = "3",
    year = "2026",
    journal = "",
}

@article{ET:2025xjr,
    author = "Abac, Adrian and others",
    collaboration = "ET",
    title = "{The Science of the Einstein Telescope}",
    eprint = "2503.12263",
    archivePrefix = "arXiv",
    primaryClass = "gr-qc",
    reportNumber = "ET-0036C-25",
    month = "3",
    year = "2025",
    journal = "",
}

@article{LISA:2024hlh,
    author = "Colpi, Monica and others",
    collaboration = "LISA",
    title = "{LISA Definition Study Report}",
    eprint = "2402.07571",
    archivePrefix = "arXiv",
    primaryClass = "astro-ph.CO",
    month = "2",
    year = "2024",
    journal = "",
}

@article{Mitman:2021xkq,
    author = "Mitman, Keefe and others",
    title = "{Fixing the BMS frame of numerical relativity waveforms}",
    eprint = "2105.02300",
    archivePrefix = "arXiv",
    primaryClass = "gr-qc",
    doi = "10.1103/PhysRevD.104.024051",
    journal = "Phys. Rev. D",
    volume = "104",
    number = "2",
    pages = "024051",
    year = "2021"
}

@article{Mitman:2022kwt,
    author = "Mitman, Keefe and others",
    title = "{Fixing the BMS frame of numerical relativity waveforms with BMS charges}",
    eprint = "2208.04356",
    archivePrefix = "arXiv",
    primaryClass = "gr-qc",
    doi = "10.1103/PhysRevD.106.084029",
    journal = "Phys. Rev. D",
    volume = "106",
    number = "8",
    pages = "084029",
    year = "2022"
}

@article{Buonanno:2010yk,
    author = "Buonanno, Alessandra and Kidder, Lawrence E. and Mroue, Abdul H. and Pfeiffer, Harald P. and Taracchini, Andrea",
    title = "{Reducing orbital eccentricity of precessing black-hole binaries}",
    eprint = "1012.1549",
    archivePrefix = "arXiv",
    primaryClass = "gr-qc",
    doi = "10.1103/PhysRevD.83.104034",
    journal = "Phys. Rev. D",
    volume = "83",
    pages = "104034",
    year = "2011"
}

@article{Corman:2025wun,
    author = "Corman, Maxence and Arest{\'e} Sal{\'o}, Llibert and Clough, Katy",
    title = "{Black hole binaries in shift-symmetric Einstein-scalar-Gauss-Bonnet gravity experience a slower merger phase}",
    eprint = "2511.19073v1",
    archivePrefix = "arXiv",
    primaryClass = "gr-qc",
    month = "11",
    year = "2025",
    journal = "",
}

@article{AresteSalo:2025sxc,
    author = "Arest{\'e} Sal{\'o}, Llibert and Doneva, Daniela D. and Clough, Katy and Figueras, Pau and Yazadjiev, Stoytcho S.",
    title = "{Challenges in the nonlinear evolution of unequal mass binaries in scalar-Gauss-Bonnet gravity}",
    eprint = "2507.13046",
    archivePrefix = "arXiv",
    primaryClass = "gr-qc",
    doi = "10.1103/tr7v-jhhm",
    journal = "Phys. Rev. D",
    volume = "112",
    number = "8",
    pages = "084022",
    year = "2025"
}

@article{Coates:2023swo,
    author = "Coates, Andrew and Ramazano{\u{g}}lu, Fethi M.",
    title = "{Treatments and placebos for the pathologies of effective field theories}",
    eprint = "2307.07743",
    archivePrefix = "arXiv",
    primaryClass = "gr-qc",
    doi = "10.1103/PhysRevD.108.L101501",
    journal = "Phys. Rev. D",
    volume = "108",
    number = "10",
    pages = "L101501",
    year = "2023"
}

@article{Rubio:2024ryv,
    author = "Rubio, Marcelo E. and Lara, Guillermo and Bezares, Miguel and Crisostomi, Marco and Barausse, Enrico",
    title = "{Fixing the dynamical evolution of self-interacting vector fields}",
    eprint = "2407.08774",
    archivePrefix = "arXiv",
    primaryClass = "gr-qc",
    doi = "10.1103/PhysRevD.110.063015",
    journal = "Phys. Rev. D",
    volume = "110",
    number = "6",
    pages = "063015",
    year = "2024"
}

@article{Bezares:2021yek,
    author = "Bezares, Miguel and ter Haar, Lotte and Crisostomi, Marco and Barausse, Enrico and Palenzuela, Carlos",
    title = "{Kinetic screening in nonlinear stellar oscillations and gravitational collapse}",
    eprint = "2105.13992",
    archivePrefix = "arXiv",
    primaryClass = "gr-qc",
    doi = "10.1103/PhysRevD.104.044022",
    journal = "Phys. Rev. D",
    volume = "104",
    number = "4",
    pages = "044022",
    year = "2021"
}

@article{Gasparotto:2026bru,
    author = "Gasparotto, Silvia and Zosso, Jann and Arest{\'e} Sal{\'o}, Llibert and Doneva, Daniela D. and Yazadjiev, Stoytcho S.",
    title = "{Gravitational Memory from Hairy Binary Black Hole Mergers}",
    eprint = "2604.09350",
    archivePrefix = "arXiv",
    primaryClass = "gr-qc",
    reportNumber = "CERN-TH-2026-086",
    month = "4",
    year = "2026",
    journal = "",
}

@article{Hannam:2009hh,
    author = "Hannam, Mark and others",
    title = "{The Samurai Project: Verifying the consistency of black-hole-binary waveforms for gravitational-wave detection}",
    eprint = "0901.2437",
    archivePrefix = "arXiv",
    primaryClass = "gr-qc",
    doi = "10.1103/PhysRevD.79.084025",
    journal = "Phys. Rev. D",
    volume = "79",
    pages = "084025",
    year = "2009"
}

@article{Aylott:2009ya,
    author = "Aylott, Benjamin and others",
    title = "{Testing gravitational-wave searches with numerical relativity waveforms: Results from the first Numerical INJection Analysis (NINJA) project}",
    eprint = "0901.4399",
    archivePrefix = "arXiv",
    primaryClass = "gr-qc",
    doi = "10.1088/0264-9381/26/16/165008",
    journal = "Class. Quant. Grav.",
    volume = "26",
    pages = "165008",
    year = "2009"
}

@article{Aylott:2009tn,
    author = "Aylott, Benjamin and others",
    editor = "Sutton, Patrick and Shoemaker, Deirdre",
    title = "{Status of NINJA: The Numerical INJection Analysis project}",
    eprint = "0905.4227",
    archivePrefix = "arXiv",
    primaryClass = "gr-qc",
    doi = "10.1088/0264-9381/26/11/114008",
    journal = "Class. Quant. Grav.",
    volume = "26",
    pages = "114008",
    year = "2009"
}

@article{Ripley:2022cdh,
    author = "Ripley, Justin L.",
    title = "{Numerical relativity for Horndeski gravity}",
    eprint = "2207.13074",
    archivePrefix = "arXiv",
    primaryClass = "gr-qc",
    doi = "10.1142/S0218271822300178",
    journal = "Int. J. Mod. Phys. D",
    volume = "31",
    number = "13",
    pages = "2230017",
    year = "2022"
}

@article{Julie:2024fwy,
    author = "Juli{\'e}, F{\'e}lix-Louis and Pompili, Lorenzo and Buonanno, Alessandra",
    title = "{Inspiral-merger-ringdown waveforms in Einstein-scalar-Gauss-Bonnet gravity within the effective-one-body formalism}",
    eprint = "2406.13654",
    archivePrefix = "arXiv",
    primaryClass = "gr-qc",
    doi = "10.1103/PhysRevD.111.024016",
    journal = "Phys. Rev. D",
    volume = "111",
    number = "2",
    pages = "024016",
    year = "2025"
}

@article{Allwright:2018rut,
    author = "Allwright, Gwyneth and Lehner, Luis",
    title = "{Towards the nonlinear regime in extensions to GR: assessing possible options}",
    eprint = "1808.07897",
    archivePrefix = "arXiv",
    primaryClass = "gr-qc",
    doi = "10.1088/1361-6382/ab0ee1",
    journal = "Class. Quant. Grav.",
    volume = "36",
    number = "8",
    pages = "084001",
    year = "2019"
}

@software{deppe_2026_19373346,
  author       = {Deppe, Nils and
                  Throwe, William and
                  Kidder, Lawrence E. and
                  Vu, Nils L. and
                  Nelli, Kyle C. and
                  Armaza, Cristóbal and
                  Bonilla, Marceline S. and
                  Hébert, François and
                  Kim, Yoonsoo and
                  Kumar, Prayush and
                  Lovelace, Geoffrey and
                  Macedo, Alexandra and
                  Moxon, Jordan and
                  O'Shea, Eamonn and
                  Pfeiffer, Harald P. and
                  Scheel, Mark A. and
                  Teukolsky, Saul A. and
                  Wittek, Nikolas A. and
                  Anantpurkar, Isha and
                  Anderson, Carter and
                  Boyle, Michael and
                  Carpenter, Alexander and
                  Ceja, Andrea and
                  Chaudhary, Himanshu and
                  Corso, Nicholas and
                  Dittmer, Clemens and
                  Fayyazuddin Ljungberg, Nora and
                  Foucart, Francois and
                  Ghadiri, Noora and
                  Giesler, Matthew and
                  Guo, Jason S. and
                  Habib, Sarah and
                  Huang, Chenhang and
                  Iozzo, Dante A. B. and
                  Jones, Ken Z. and
                  Lara, Guillermo and
                  Legred, Isaac and
                  Li, Dongjun and
                  Ma, Sizheng and
                  Melchor, Denyz and
                  Mendes, Iago and
                  Morales, Marlo and
                  Most, Elias R. and
                  Murphy, Michael and
                  Nee, Peter James and
                  Nishimura, Nami and
                  Osorio, Alejandro and
                  Pajkos, Michael A. and
                  Pannone, Kyle and
                  Pineda, Jose Maria and
                  Prasad, Vaishak and
                  Ramirez, Teresita and
                  Ring, Noah and
                  Rüter, Hannes R. and
                  Sanchez, Jennifer and
                  Stein, Leo C. and
                  Tellez, Daniel and
                  Thomas, Sierra and
                  Tommasini, Vittoria and
                  Vieira, Daniel and
                  Wang, Xiyue and
                  Wlodarczyk, Tom and
                  Wu, David and
                  Yoo, Jooheon},
  title        = {SpECTRE},
  month        = apr,
  year         = 2026,
  publisher    = {Zenodo},
  version      = {2026.04.01},
  doi          = {10.5281/zenodo.19373346},
  url          = {https://doi.org/10.5281/zenodo.19373346},
  swhid        = {swh:1:dir:2f4d75ce1697988e65aa07d19d8a99a9cc838db9
                   ;origin=https://doi.org/10.5281/zenodo.4290404;vis
                   it=swh:1:snp:f0e0453d78e7032932dda78e4a9638605a8fa
                   ef0;anchor=swh:1:rel:49a4961c95dcc0e30fc3c1e4c302f
                   210d873a340;path=sxs-collaboration-spectre-f6822a2
                  },
}

@article{Szilagyi:2009qz,
    author = "Szilagyi, Bela and Lindblom, Lee and Scheel, Mark A.",
    title = "{Simulations of Binary Black Hole Mergers Using Spectral Methods}",
    eprint = "0909.3557",
    archivePrefix = "arXiv",
    primaryClass = "gr-qc",
    doi = "10.1103/PhysRevD.80.124010",
    journal = "Phys. Rev. D",
    volume = "80",
    pages = "124010",
    year = "2009"
}

@article{Deppe:2018uye,
    author = "Deppe, Nils and Kidder, Lawrence E. and Scheel, Mark A. and Teukolsky, Saul A.",
    title = "{Critical behavior in 3D gravitational collapse of massless scalar fields}",
    eprint = "1802.08682",
    archivePrefix = "arXiv",
    primaryClass = "gr-qc",
    doi = "10.1103/PhysRevD.99.024018",
    journal = "Phys. Rev. D",
    volume = "99",
    number = "2",
    pages = "024018",
    year = "2019"
}

@article{Choptuik:2009ww,
    author = "Choptuik, Matthew W. and Pretorius, Frans",
    title = "{Ultra Relativistic Particle Collisions}",
    eprint = "0908.1780",
    archivePrefix = "arXiv",
    primaryClass = "gr-qc",
    doi = "10.1103/PhysRevLett.104.111101",
    journal = "Phys. Rev. Lett.",
    volume = "104",
    pages = "111101",
    year = "2010"
}

@article{Campanelli:2005dd,
    author = "Campanelli, Manuela and Lousto, C. O. and Marronetti, P. and Zlochower, Y.",
    title = "{Accurate evolutions of orbiting black-hole binaries without excision}",
    eprint = "gr-qc/0511048",
    archivePrefix = "arXiv",
    doi = "10.1103/PhysRevLett.96.111101",
    journal = "Phys. Rev. Lett.",
    volume = "96",
    pages = "111101",
    year = "2006"
}

@article{Sanger:2024axs,
    author = {S{\"a}nger, Elise M. and others},
    title = "{Tests of General Relativity with GW230529: a neutron star merging with a lower mass-gap compact object}",
    eprint = "2406.03568",
    archivePrefix = "arXiv",
    primaryClass = "gr-qc",
    reportNumber = "LIGO-P2400200",
    month = "6",
    year = "2024",
    journal = "",
}

@article{LIGOScientific:2026uyd,
    collaboration = "LIGO Scientific, VIRGO, KAGRA",
    title = "{GWTC-5.0: Constraints on the Cosmic Expansion Rate and Modified Gravitational-wave Propagation}",
    eprint = "2605.27227",
    archivePrefix = "arXiv",
    primaryClass = "astro-ph.CO",
    reportNumber = "LIGO-P2600018",
    month = "5",
    year = "2026",
    journal = "",
}

@article{Cardoso:2020iji,
    author = "Cardoso, Vitor and Macedo, Caio F. B. and Vicente, Rodrigo",
    title = "{Eccentricity evolution of compact binaries and applications to gravitational-wave physics}",
    eprint = "2010.15151",
    archivePrefix = "arXiv",
    primaryClass = "gr-qc",
    doi = "10.1103/PhysRevD.103.023015",
    journal = "Phys. Rev. D",
    volume = "103",
    number = "2",
    pages = "023015",
    year = "2021"
}

@misc{SXSPackage_v2025.0.25,
    doi = {10.5281/ZENODO.18951276},
    url = {https://zenodo.org/doi/10.5281/zenodo.18951276},
    author = {Boyle, Michael and Mitman, Keefe and Scheel, Mark and Stein, Leo},
    title = {The sxs package},
    publisher = {Zenodo},
    year = {2026},
    copyright = {MIT License}
    }

@article{Alcubierre:2003pc,
    author = "Alcubierre, Miguel and others",
    title = "{Toward standard testbeds for numerical relativity}",
    eprint = "gr-qc/0305023",
    archivePrefix = "arXiv",
    reportNumber = "AEI-2003-041",
    doi = "10.1088/0264-9381/21/2/019",
    journal = "Class. Quant. Grav.",
    volume = "21",
    number = "2",
    pages = "589--613",
    year = "2004"
}

%


%
%

\end{document}